\documentstyle[twoside,fleqn,espcrc2]{article}

\newcommand{\AmS}{{\protect\the\textfont2
  A\kern-.1667em\lower.5ex\hbox{M}\kern-.125emS}}

\newcommand{\beq}{\begin{equation}}
\newcommand{\eeq}{\end{equation}}
\newcommand{\beqs}{\begin{eqnarray}}
\newcommand{\eeqs}{\end{eqnarray}}

\title{Complex-Temperature Singularities of Ising Models}

\author{Robert Shrock\address{Institute for Theoretical Physics,
        State University of New York,
        Stony Brook, NY  11794 USA}%
        \thanks{email: shrock@max.physics.sunysb.edu.  Research partially
        supported by the U.S. NSF under grant PHY-93-09888.}}

\begin{document}

\begin{abstract}
We report new results on complex-temperature properties of Ising models.
These include studies of the $s=1/2$ model on triangular, honeycomb,
kagom\'e, $3 \cdot 12^2$, and $4 \cdot 8^2$ lattices.  We elucidate the
complex--$T$ phase diagrams of the higher-spin 2D Ising models, using
calculations of partition function zeros.
Finally, we investigate the 2D Ising model in an external magnetic field,
mapping the complex--$T$
phase diagram and exploring various singularities therein.
For the case $\beta H=i\pi/2$, we give exact results on the phase
diagram and obtain susceptibility exponents $\gamma'$ at various
singularities from low-temperature series analyses.

\end{abstract}

\maketitle

\section{Introduction}

   During the past year, in collaboration with Victor Matveev, we have
carried out a number of studies of complex-temperature (CT) properties of
Ising models \cite{chisq}-\cite{only}.  We report on some results from
these here (see also \cite{ms,lat94}).
There are several reasons for studying the properties of statistical
mechanical models with the temperature variable generalized to take on complex
values.  First, one can understand more deeply the behavior of various
thermodynamic quantities by seeing how they behave as analytic functions of
complex temperature.  Indeed, complex--$T$ singularities have important
influences on the behavior of such functions on the real $K$ axis (see
definition in
eq. (\ref{kh})). Second, one can see how the physical phases of a given
model generalize to regions in appropriate complex-temperature variables.
Third, a knowledge of the complex-temperature singularities of quantities
which have not been calculated exactly helps in the search for exact,
closed-form expressions for these quantities.  This applies, in particular,
to the susceptibility of the 2D zero-field Ising model, and to the Ising model
for higher spin or external magnetic field, neither of which has been solved
above $d=1$.  Some of the earliest works on complex--$T$ properties of the
Ising model include Refs. \cite{fisher}-\cite{dg}.  For comparison, a different
direction is to study the case in which the temperature is kept physical but
the external field $H$ is complex \cite{ly}.  We also report new results which
both $K$ and $H$ are complex.

\section{Model}

   The Ising model, and our notation, are specified as follows: the partition
function is
$Z = \sum_{\{\sigma_i\}} e^{-\beta {\cal H}}$ with
\beq
{\cal H} = -J \sum_{<ij>} \sigma_i \sigma_j - H \sum_i \sigma_i
\label{h}
\eeq
where $\sigma_i = \pm 1$, $\beta = (k_BT)^{-1}$, and $H$ denotes a possible
external field (zero unless otherwise indicated). We use the
standard notation
\beq
K = \beta J \ , \quad h = \beta H
\label{kh}
\eeq
\beq
z = e^{-2K} \ , \quad u=e^{-4K} \ , \quad \mu=e^{-2h}
\label{zumu}
\eeq
$v = \tanh K$, and the elliptic modulus variable
$k_<=4u/(1-u)^2$.  We define the reduced free energy as $f=-\beta F = \lim_{N
\to \infty} N^{-1} \ln Z$ and the
reduced susceptibility as $\bar\chi=\beta^{-1}\chi$.

\section{Phase Diagrams (Homopolygonal Lattices)}

     Owing to the
symmetries under the shifts $K \to K + n i \pi$ and $K \to K + (2n+1)(i\pi/2)$
\cite{ms}, there is an infinite replication of complex phases as functions of
$K$; this is reduced to a single set by using the variables
$v$ or $z$.  For a bipartite lattice, the phase boundaries are invariant
under $z \to -z$, so that the most compact way to show them is in terms of the
variable $u=z^2$.  In Ref. \cite{chisq} we
showed that for the square lattice the phase boundaries in the $u$ plane are
given by a Pascal lima\c{c}on.  The inner and outer branches of this
lima\c{c}on cross at the point $u=-1$.

\vspace{2mm}

   We have extended this analysis to the triangular and honeycomb lattices
\cite{chitri}.  As in the case of the square lattice, the continuous locus of
points where $f$ is non-analytic can be determined exactly. ($f$ is also
trivially non-analytic at the points $|K|=\infty$, but since these do not
separate any phases, they will not be important here.)  In contrast to the
square lattice, this (continuous)
locus of points includes arcs or line segments which
protrude into, and terminate within, certain phases and hence do not completely
separate any phases.  For example, for the triangular lattice, in the $u$
plane, the above locus of points consists of the union of the circle
$u = (-1+2e^{i\theta})/3$
and the semi-infinite line segment $-\infty \le u \le -1/3$.  The interior
(exterior) of the above circle forms the FM (PM) phase.  $M$ diverges at
$u=-1/3$ with exponent $\beta=-1/8$
and is discontinuous across the line segment $-1 < u < -1/3$.
We have proved that such a divergence in $M$ automatically
implies a divergence in $\bar\chi$ \cite{chitri}.

\section{Heteropolygonal Lattices}

   While the square, triangular, and honeycomb lattices each involve only one
type of regular polygon (and thus may be denoted ``homopolygonal''),
there are also regular tilings of the plane (heteropolygonal lattices) which
involve more than one type of regular polygons, but still have all vertices
equivalent, and all bonds of equal length.  We have
determined the exact complex-temperature phase diagrams of
the 2D Ising model on three regular heteropolygonal lattices:
$3 \cdot 6 \cdot 3 \cdot 6$ ($=$ kagom\'{e}), $3 \cdot 12^2$, and
$4 \cdot 8^2$ (bathroom tile), where the notation denotes the regular
$n$-sided polygons adjacent to each vertex \cite{cmo}.
We have also worked out the exact complex--$T$ singularities of $M$.
In particular, we found the first case (the $4 \cdot 8^2$ lattice)
where, even for
isotropic spin-spin exchange couplings, the nontrivial non-analyticities of
the free energy of the Ising model lie in a two-dimensional, rather than
one-dimensional, algebraic variety in the $z=e^{-2K}$ plane.

\section{Complex--$T$ Singularities}

   Complex--$T$ singularities exhibit a number of new features not observed for
physical critical points.
First, as we showed by exact results \cite{chisq}, universality, in the sense
of lattice-independence of critical exponents, is violated at
$u=-1$, since the magnetization exponent $\beta$ is 1/4 for the square lattice
but 3/8 for the triangular and honeycomb lattices.  Second, on the square
lattice, we proved \cite{ms} that as one approaches the boundary of the
complex--$T$ extension of the PM phase, $\bar\chi$ has at most finite
non-analyticities except for its divergence at the physical critical point
$u_c$, and hence, in particular, at $u=-1\equiv u_s$, it follows that
$\gamma_s < 0$.  Analyses of low-temperature series indicate that $\bar\chi$
diverges at $u=-1$ as this point is approached from within the FM phase
\cite{egj,chisq} (and AFM phase \cite{chisq}), with the exponent $\gamma_s'$
numerically consistent with 3/2.  This value also follows from the relation
$\gamma_s'=2(\gamma'-1)$ \cite{chisq}.  Hence, $\gamma_s \ne \gamma_s'$, which
is unprecedented for physical critical points.
Third, the correlation length exponent $\nu_s'$ has different values
depending on whether one extracts it from non-diagonal correlation functions
or diagonal correlation functions.
Several scaling relations are violated; for example, $\alpha_s + 2\beta_s +
\gamma_s \ne 2$ for the singularity at $u=-1$ as approached
from within the PM phase.  For honeycomb lattice, we have shown
\cite{chitri} that $\alpha' + 2\beta + \gamma' \ne 2$ (the RHS is consistent
with being equal to 4) at the CT singularity at $z=-1$, as approached from
within the FM phase, and have discussed the origin of this.

\section{Higher-Spin Ising Model}

 We have carried out a study of the complex--$T$ properties of
higher-spin 2D Ising models \cite{hs}.  The method which we use for this
purpose is the calculation of complex--$T$ zeros of the partition function
on finite lattices.  It is known from exactly solved cases
that as the lattice size increases, the CT zeros of $Z$ occur on, or
progressively closer to, the continuous locus of points where $f$ is
non-analytic in the thermodynamic limit.  Hence, by studying these zeros, one
can make reasonable inferences about this locus of points and the corresponding
complex-$T$ phase diagram.  Specific results are presented for $s=1$, 3/2, and
2 in Ref. \cite{hs}.
We infer several general features of the
phase diagrams and motivate a conjecture that the number of arcs protruding
into the complex-temperature extension of the FM phase is $4[s^2]-2$ for $s \ge
1$, where $[x]$ denotes the integral part of $x$.  From comparison with an
analysis of low-temperature series expansions \cite{jge}, further support is
obtained for our conjecture that the magnetization diverges at endpoints of
such arcs terminating in the FM phase.  We also present an exact
determination of the complex-temperature phase diagram for the 1D spin $s$
Ising model.  We find that in the complex $u_s=e^{-K/s^2}$ plane
(i) it consists of $N_{c,1D}=4s^2$ infinite regions separated by an equal
number of boundary curves where the free energy is non-analytic;
(ii) these curves extend from the origin to complex
infinity, and in both limits are oriented along the angles $\theta_n =
(1+2n)\pi/(4s^2)$, for $n=0,..., 4s^2-1$; (iii)
there is a boundary curve (line) along the negative real
$u_s$ axis if and only if $s$ is half-integral.
An interesting connection is noted between some features of the
1D and 2D complex-temperature phase diagrams for higher-spin Ising models.

\section{Finite Magnetic Field}

   Although the Ising model has never been solved in an arbitrary
external magnetic field, Lee and Yang \cite{ly} did succeed in solving
exactly for the free energy and magnetization for the model on
the square lattice for $\beta H \equiv h=i \pi/2$.  This solution yields
further insight into the structure of the model.  It is interesting to
investigate the properties of this model when one also generalizes the
temperature to complex values, and we have done this \cite{ih}.
We have first determined the continuous locus of points where $f$ is
non-analytic for this model: in the $u$ plane this is the union of
(i) the unit circle and (ii) the finite line segment along the negative real
axis
\beq
1/u_e \le u \le u_e
\label{ulinesegment}
\eeq
where
\beq
u_e = -(3 - 2^{-3/2})
\label{ue}
\eeq
The complex--$T$ phase diagram of this model
consists of a FM phase inside of, and an AFM phase outside of,
the unit circle in the $u$ plane.
We find that the specific heat $C$ has the following singularities:
divergences, with exponent $\alpha_e'=1$ at $u_e$ and $1/u_e$, and finite
logarithmic non-analyticities at $u=-1\equiv u_s$ and $u=1$ ($\alpha_s'=0$,
$\alpha_1'=0$).  The uniform and staggered
magnetizations $M$ and $M_{st}$ vanish
with exponent $\beta_s=1/2$ at $u=-1$ and diverge at $u=1$ with
exponent $\beta_1=-1/4$.  $M$ and $M_{st}$ diverge, respectively, at $u_e$ and
$1/u_e$, with exponent $\beta_e=-1/8$.

   From the Baxter-Enting
low-temperature, high-field series expansion of the partition function, we
have extracted the low-temperature series
for the susceptibility $\bar \chi$ to $O(u^{23})$.  Analyzing this series, we
find that $\bar\chi$ has divergent singularities (i)
at $u=u_e$ with exponent $\gamma_e'=1.25 \pm 0.01$; (ii) at $u=1$, with
exponent $\gamma_1'=2.50 \pm 0.01$; and (iii) at $u=u_s=-1$, with exponent
$\gamma_s'=1.00 \pm 0.08$. We infer the exact values of these exponents to be
\beq
\gamma_e' = \frac{5}{4}
\label{gammae}
\eeq
\beq
\gamma_1' = \frac{5}{2}
\label{gamma1}
\eeq
\beq
\gamma_s' = 1
\label{gammas}
\eeq

   We have subsequently determined the exact complex--$T$
phase diagram for the model at
$h=i\pi/2$ on the honeycomb ($hc$) and triangular ($t$)
lattices \cite{only}.  For the $t$ lattice, we find that this consists of a
single (FM) phase, and the continuous locus of points where $f$ is non-analytic
consists of the union of (i) the circular arc
\beq
u = \frac{1}{2}(-1 + e^{i\theta}) \ , \quad  \theta_{ce} \le |\theta| \le \pi
\label{ucirclem1}
\eeq
where $\theta_{ce}=arctan(4\sqrt{2}/7) \simeq 38.9^\circ$,
corresponding to the endpoints $u_{ce}=e^{i\theta_{ce}}$ and $u_{ce}^*$, where
$u_{ce}=(-1 + 2^{3/2}i)/9$, and (ii) the semi-infinite line segment
\beq
 -\infty \le u \le -1/2
\label{usegmentm1}
\eeq
For the $hc$ lattice, we have proved that all of our results on the
complex--$T$ phase diagram and the singularities of
$\bar\chi$ for $h=0$ on the $hc$ lattice also apply for $h=i\pi/2$ with the
replacement of $z$ by $-z$.

\vspace{3mm}

   In Ref. \cite{only}
we have recently investigated the complex--$T$ phase diagram of
the square-lattice Ising model for nonzero external magnetic field
$H$ in the range $-\infty \le H \le \infty$, i.e. $0 \le \mu \le \infty$ where
$\mu$ is given by (\ref{zumu}).
We have also carried out a similar analysis for $-\infty \le \mu \le 0$.
Without loss of generality, one may restrict to the range $-1 \le \mu \le 1$.
The results for this interval
provide a new way of continuously connecting the two known exact solutions of
this model, viz., that at $\mu=1$ ($h=0$) and $\mu=-1$ ($h=i\pi/2$).  Our
methods include calculations of complex-temperature zeros of the partition
function and analysis of low-temperature series expansions.  For real nonzero
$H$, we find that
the inner branch of a lima\c{c}on bounding the FM phase breaks and forms
two complex-conjugate arcs which move away from the real $u$ axis as $|H|$
increases.  We have studied the singularities and associated
exponents of thermodynamic functions at the endpoints of these arcs.  For $\mu
< 0$, there are two line segments of singularities on the negative and positive
$u$ axis, and we carry out a similar study of the behavior at the inner
endpoints of these arcs, which constitute the nearest singularities to the
origin in this case.

\section{Conclusions}

   Thus, the study of complex-temperature singularities of the Ising model
provides many interesting insights.  A number of topics deserve further
investigation, such as related spin models and behavior in $d > 2$.

\end{document}